\PassOptionsToPackage{colorlinks=true,allcolors=black}{hyperref}

\documentclass[11pt]{article}

\usepackage[T1]{fontenc}
\usepackage{lmodern}

\usepackage{setspace}
\usepackage{caption}
\usepackage{amsmath, amssymb}
\usepackage{array}
\usepackage[table]{xcolor}
\usepackage{colortbl}
\usepackage{booktabs}
\usepackage{graphicx}

\usepackage{amsthm}

\newcommand{\PRS}{\textsf{PRS}}
\newcommand{\BRK}{\textsf{BRK}}
\newcommand{\NEU}{\textsf{NEU}}
\newcommand{\MIX}{\textsf{MIX}}
\newcommand{\NA}{\textsf{N/A}}

\newcommand{\OBL}{\mathrm{OBL}}
\newcommand{\OCC}{\mathrm{OCC}}
\newcommand{\REC}{\mathrm{REC}}

\newcommand{\LOC}{\mathrm{LOC}}
\newcommand{\OBJ}{\mathrm{OBJ}}

\newcommand{\SCOPEE}{\mathrm{SCOPE\text{-}E}}
\newcommand{\SCOPES}{\mathrm{SCOPE\text{-}S}}

\newcommand{\RULEC}{\mathrm{RULE\text{-}C}}
\newcommand{\RULES}{\mathrm{RULE\text{-}S}}

\newcommand{\Fset}{\mathcal{F}}
\newcommand{\Class}{\mathrm{Class}}

\newcommand{\DefRef}[2]{\hyperref[#1]{Definition (#2)}}
\newcommand{\ConstRef}[2]{\hyperref[#1]{Constraint (#2)}}

\newcommand{\Substrate}{\mathcal{S}}
\newcommand{\SubstrateRef}{\mathcal{S}_{\mathrm{ref}}}
\newcommand{\Frameworks}{\mathbb{F}}

\theoremstyle{plain}
\newtheorem{theorem}{Theorem}[section]
\newtheorem{lemma}[theorem]{Lemma}
\newtheorem{proposition}[theorem]{Proposition}

\theoremstyle{definition}
\newtheorem{definition}[theorem]{Definition}

\theoremstyle{remark}

\newtheorem*{theorem*}{Theorem}
\newtheorem*{lemma*}{Lemma}
\newtheorem*{proposition*}{Proposition}
\newtheorem*{corollary*}{Corollary}
\newtheorem*{conjecture*}{Conjecture}
\newtheorem*{claim*}{Claim}
\newtheorem*{definition*}{Definition}
\newtheorem*{example*}{Example}
\newtheorem*{remark*}{Remark}
\newtheorem*{note*}{Note}
\newtheorem*{convention*}{Convention}

\usepackage{url}
\usepackage{natbib}
\bibpunct{(}{)}{;}{a}{}{ }

\usepackage{hyperref}
\usepackage{bookmark}

\usepackage{authblk}

\usepackage{microtype}

\providecommand{\keywords}[1]{\textbf{Keywords: } #1}

\usepackage{enumitem}
\setlist[itemize]{itemsep=0.2em, topsep=0.2em, parsep=0em, partopsep=0em}

\hypersetup{
  pdftitle={Referential Regimes: Transformation-Based Identity Profiles for Neutral Substrates},
  pdfauthor={Denise M. Case},
  pdfkeywords={formal ontology, identity and persistence, neutral substrates, accountability}
}

\title{Referential Regimes:  \\
Transformation-Based Identity Profiles \\
for Neutral Substrates}
\author{Denise M. Case}
\affil{Northwest Missouri State University, Computer Science and Information Systems, Maryville, MO, USA}
\date{} 

\begin{document}

\maketitle
\vspace{-1em}

\begin{abstract}
  Data systems operating under persistent legal, institutional,
  and analytic disagreement require shared reference without
  presupposing agreement about interpretation.
  This paper investigates identity preservation under
  transformation, building on the conditional neutrality
  result of \emph{Neutral Substrates}.

  We pair candidate identity bases with families of record
  transformations and classify each family as preserving, breaking,
  neutral, or inapplicable for a basis.
  Using invariant-defined identity bases and explicit effect
  models, we derive and certify six classification cells in Lean~4
  for scope decomposition, rule refinement, and forking.
  The derivations show how classifications depend on field
  granularity, admissible transformation steps, and explicit
  continuity warrants.

  We also characterize the identity relations generated
  by invariant-preserving steps.
  Two bases generate equal relations exactly when they agree on
  the preservation of every admissible step in a shared comparison
  domain, and a single separating step witnesses a difference.

  The formalization makes identity claims auditable, provides
  testable conditions for systems that transform records, and
  exposes the assumptions behind each classification.
  The results are conditional on the declared bases, models, and
  warrants, whose adequacy for particular referents requires
  independent justification.
\end{abstract}



\section{Introduction}
\label{sec:intro}

Modern accountability systems increasingly operate under legal,
institutional, and analytic disagreement.
A shared record may need to
support reference to the same intended subject while its representations
are amended, forked, refined, decomposed, aggregated, or reclassified.
Agreement about what a record means cannot be assumed, and neither can
agreement about which identity conditions apply to its referents.
This paper investigates candidate ways to characterize individuation,
co-reference, and persistence under those transformations.

\emph{Neutral Substrates}~\citep{case2026neutral} provides the motivating
constraint: under Referential Common Ground and the stated foundational-layer
restriction, a substrate satisfies semantic neutrality relative to its
admissible frameworks.
That conditional result does not establish a
referential identity criterion for a particular bearer.
The present paper
examines what a declared criterion and selected transformation behavior
can express, while keeping the adequacy of those choices open to
independent justification.

The analysis pairs proposed identity bases with classifications of selected
transformation families.
An identity regime, in the terminology of this
paper, is a declared basis together with its transformation profile.
A classification records a proposed preservation, breakage, neutrality,
or inapplicability judgment relative to a declared basis.
Relations generated from admissible transformation steps are formal
constructions; their interpretation as strict identity requires
independent justification.

We select reference kinds for analysis: obligation-bearers, rules,
occurrences, scopes, records, and plain referents.
For each we propose candidate carrier--basis profiles.
Some kinds receive paired bases that distinguish a locus
from an object occupying it, a scope's coverage from its internal
structure, and a rule's content from its structure.
Forking, aggregation or decomposition, and refinement provide
tests of those distinctions.

The candidate profiles have pairwise distinct classification
functions, and the carrier--basis assignment is determinate.
Whether distinct profiles generate distinct equivalence relations
depends on the generating rule, the admissible transformation
dynamics, and the comparison domain (Appendix~\ref{app:sep}).
A carrier-indexed combination holds separate relations within one
explicitly tagged relation (Section~\ref{sec:distinctions}).

The principal contribution is a formal method for
deriving and examining identity-preservation classifications
under transformation.

First, identity claims become auditable: a verdict follows
from declared identity invariants, admissible transformation
steps, and explicit assumptions that can be independently
examined.

Second, the construction supplies testable conditions for
record-transforming systems.
An admissible transformation step that preserves
one declared invariant but changes another
provides a concrete witness of different generated
identity relations.

Third, the formalization identifies which assumptions
carry each result.
The scope and rule examples expose
field-granularity and transformation-domain dependencies;
the fork example exposes the need for explicit
continuity warrants.

The classifications, generated relations, and the distinctions among
them are formal results about the declared bases and models.
Applying them to a particular referent requires an independently
justified identity criterion (Section~\ref{sec:limits}).

The remainder of the paper proceeds as follows.
Section~\ref{sec:background} states the Neutral Substrate foundation
and the narrower contribution pursued here.
Section~\ref{sec:formal_setting} sets out the formal setting and
transformation classifications.
Section~\ref{sec:inventory} specifies the selected reference inventory.
Section~\ref{sec:algebra} develops conditional comparisons among identity
bases, and Section~\ref{sec:profiles} collects the candidate profiles.
Section~\ref{sec:distinctions} compares their generated relations and
shows how carrier-specific relations combine.
Section~\ref{sec:determinacy} establishes determinacy of the selected
profile assignment, while Section~\ref{sec:application} treats
application settings as identity-criterion tests.
Section~\ref{sec:related} situates the work in prior research;
Section~\ref{sec:limits} states its boundaries; and
Section~\ref{sec:conclusion} concludes.


\section{Background: Neutral Substrates}
\label{sec:background}

This paper builds on \emph{Neutral Substrates}~\citep{case2026neutral}.
That work studies a shared representational base intended to support
reference under persistent interpretive disagreement without adopting
contested causal or normative interpretations as unqualified foundational
commitments.
The present paper investigates proposed identity conditions
and their behavior when representations are transformed.
It does not assume that the abstract referential interfaces
have been given adequate identity criteria for every intended kind of referent.

The definitions below retain their \emph{Neutral Substrates} labels and wording.
They establish the terminology imported for the present analysis.

\subsection{Inherited Definitions}
\label{subsec:ns-inherited-definitions}

\begin{definition}[Substrate]
  \label{se100.def.Substrate}
  A \emph{substrate} $\Substrate$ is a shared representational base providing
  stable reference for entities, occurrences, and institutional artifacts
  across a class of admissible interpretive frameworks $\Frameworks$.
  Stable reference consists of individuation, co-reference, and persistence.
\end{definition}

A substrate is the foundational layer, not the complete record system.
It can support later causal, normative, evidentiary, or institutional
interpretations without adopting those interpretations as its own
object-level commitments.
The distinction between a referent and
interpretations attributed to it is central; preserving a shared label
or descriptive record does not itself establish that the intended
referent has been fixed.

\begin{definition}[Referential Regime]
  \label{se100.def.ReferentialRegime}
  A \emph{referential regime} is the triple of
  \begin{enumerate}
    \item individuation conditions,
    \item co-reference conditions, and
    \item persistence conditions
  \end{enumerate}
  by which a substrate fixes and tracks entities, occurrences, and institutional
  artifacts.
\end{definition}

Individuation concerns what counts as one whole and sameness across
presentations; co-reference concerns whether references denote the same
referent; persistence concerns strict identity across settings and must
be distinguished from succession, derivation, or counterpart relations.
The abstract triple supplies interfaces for those questions, not their
independently warranted answers.
A relation defined by closure of
proposed preserving steps does not,
without additional grounding,
establish identity in an intended domain.

\begin{definition}[Referential Commitments]
  \label{se100.def.ReferentialCommitments}
  Let $\SubstrateRef$ denote the \emph{referential commitments} of $\Substrate$.
  These are the substrate-layer commitments fixed by its referential regime,
  including identifiers;
  the typing of entities, occurrences, and institutional artifacts;
  timestamps;
  provenance;
  and referential relations among them.
\end{definition}

Identifiers, types, timestamps, provenance, and relations may supply
materials for referential fixing.
Their mere presence does not justify
which object-language propositions are
fixed by a proposed referential regime.
That connection requires independently grounded interpretation.

\emph{Neutral Substrates} distinguishes a permitted attribution from
the proposition attributed: accepting the attribution does not require
accepting its content.
It also defines interpretive non-commitment,
extension stability, semantic neutrality, and neutrality by design.

Its central result is conditional sufficiency.
Under Referential Common
Ground, restricting foundational commitments to referential commitments
and permitted attribution propositions suffices for semantic neutrality
relative to every admissible framework.
Permitted attributions are
determined by the referential commitments.

The sufficiency result is conditional on Referential Common Ground
and the stated restriction on foundational commitments.
Its application requires independently justified referential criteria;
the present paper inherits these obligations.

\subsection{Contribution}
\label{subsec:contribution}

This paper selects a finite reference inventory, candidate identity
bases, and transformation families through which to investigate
identity-sensitive distinctions.
It studies classifications of
re-expression, annotation, refinement, aggregation or decomposition,
re-contextualization, reassignment, substitution, branching,
provenance extension, and state evolution.
The resulting classification
profiles support formal comparison of candidate identity bases.

\begin{definition}[Transformation-Invariant Identity]
  \label{se200.def.TransformationInvariantIdentity}
  Relative to a specified referent kind, proposed identity basis, and
  admissible transformation instances, a presentation is treated as
  \emph{transformation-invariant} along steps classified as preserving
  the declared basis.
  A transformation classification expresses a
  proposed identity judgment relative to that basis; it does not,
  without independent justification, establish strict identity of
  the represented referent across those steps.
\end{definition}

Section~\ref{sec:formal_setting} formalizes candidate profiles and a
relation generated from non-breaking steps;
Section~\ref{sec:algebra} compares selected basis-sensitive classifications;
Section~\ref{sec:distinctions} compares generated relations and their
combination; and Section~\ref{sec:limits} states the boundaries of the
analysis.


\section{Formal Setting}
\label{sec:formal_setting}

This paper investigates candidate conditions for stable reference under
transformation in settings of persistent interpretive disagreement.
Its formal results are relative to the selected carriers, identity bases,
transformation families, and assumptions stated below.

\subsection{Persistent Disagreement}
\label{subsec:persistent_disagreement}

We consider environments in which interpretive frameworks may remain
in disagreement and their extensions need not be coordinated.
Frameworks may advance incompatible causal accounts, normative attributions,
or interpretive classifications without being required to converge.
The neutrality constraint of Section~\ref{sec:background} concerns which
propositions the shared substrate itself adopts; it does not adjudicate
among those interpretations.

This setting assumes neither that all admissible frameworks agree nor that
they are mutually consistent.
The conditional neutrality result of
\emph{Neutral Substrates} also requires Referential Common Ground: the
complete adopted referential commitment set must remain jointly compatible
with each admissible framework.
That premise is not established by the
present inventory or transformation classifications.

Revising a substrate merely to make conflicting interpretations converge
would not resolve their disagreement neutrally.
Revising an inadequately
justified referential commitment is a different matter: any revised
construction requires renewed examination of its own compatibility and
referential warrant.
This paper does not define such a revision procedure.

\subsection{Stable Reference in Strong Form}
\label{subsec:stable_reference_strong}

We examine the demanding case in which representations of a purported
referent change through ordinary record transformations while the question
of its identity remains meaningful.
The referential regime in
\emph{Neutral Substrates} supplies abstract individuation, co-reference,
and persistence interfaces; it does not independently determine which
referents persist through a particular operation.

Stable reference requires identifying the intended referent and supplying
an adequate criterion for when presentations concern that same referent.
A change in location, role, description, or state need not constitute a
change of bearer identity; conversely, preservation of a record or label
does not establish the persistence of what it represents.
The applicable judgment depends on the proposed identity criterion and
its independent justification.

Roles, flags, contextual predicates, and carrier/basis parameters are not
inherently incompatible with neutral shared reference.
They may carry
explicit, fixed distinctions.
What matters is whether their meanings and
use are adequately specified and whether the resulting referential
commitments satisfy the relevant common-ground conditions.
A convention whose identity-relevant meaning
is left to conflicting interpretations
cannot, without further grounds, establish a shared identity judgment.

\subsection{Transformation Basis}
\label{subsec:transformation_basis}

Identity questions become visible when representations are transformed.
We select the following finite transformation basis for analysis:
\[
  \Fset =
  \{\mathrm{RE},\mathrm{AN},\mathrm{RF},\mathrm{AD},\mathrm{RC},
  \mathrm{RA},\mathrm{SU},\mathrm{BF},\mathrm{PV},\mathrm{SE}\}.
\]
The basis is a selected set of operation families, not a claim of
completeness for record systems or of identity effects uniform across
all referent kinds.

\begin{definition}[Transformation basis]
  \label{se200.def.TransformationBasis}
  The transformation basis $\Fset$ consists of these operation families:
  \begin{enumerate}[label=(\alph*),nosep]
    \item $\mathrm{RE}$, re-expression: change of representation format while preserving represented content;
    \item $\mathrm{AN}$, annotation: addition of non-identity-bearing information;
    \item $\mathrm{RF}$, refinement: increase in descriptive, structural, or normative detail;
    \item $\mathrm{AD}$, aggregation or decomposition: restructuring into coarser or finer components under a specified whole-preservation condition;
    \item $\mathrm{RC}$, re-contextualization: change in applicability context;
    \item $\mathrm{RA}$, reassignment: change in association, bearer, or assignment relation;
    \item $\mathrm{SU}$, substitution: replacement of the designated carrier;
    \item $\mathrm{BF}$, branch or fork: divergence into multiple continuations;
    \item $\mathrm{PV}$, provenance extension: addition of historical, derivational, or evidentiary metadata;
    \item $\mathrm{SE}$, state evolution: change in state of an enduring referent over time.
  \end{enumerate}
\end{definition}

These names designate typed operation families.
Their meaning and
identity effects must be evaluated relative to an intended referent,
eligible presentations, and a stated identity basis.
In particular,
$\mathrm{AD}$ may preserve the covered cases of a scope while changing
its internal organization; whether either property is identity-relevant
is a separate question.
Changes to covered cases are not represented
by that particular whole-preserving instance of $\mathrm{AD}$.

If a larger transformation basis preserves the previously assigned
classifications on $\Fset$, then any existing classification disagreement
remains a disagreement on the larger basis, as expressed in
Proposition~\ref{se200.prop.ClassificationDistinctionPersists}.

\subsection{Classification of Transformations}
\label{subsec:classification_transformations}

The construction assigns a classification to each operation family
relative to a declared identity basis and its intended comparison domain.

\begin{definition}[Transformation classification]
  \label{se200.def.TransformationClassification}
  For a profile $\tau$ and transformation family $f\in\Fset$,
  \[
    \Class_\tau(f)\in\{\PRS,\BRK,\NEU,\NA\}.
  \]
  The proposed classifications mean:
  \begin{enumerate}[label=(\alph*),nosep]
    \item $\PRS$: the operation preserves the declared identity basis;
    \item $\BRK$: the operation breaks the declared identity basis;
    \item $\NEU$: the operation is applicable but does not affect that basis;
    \item $\NA$: the operation is inapplicable on the selected domain.
  \end{enumerate}
\end{definition}

The four-valued classification above describes the stipulated
profile catalog.
The canonical Lean matrix uses
three classification values (IGN, PRS, BRK),
mapping both \NEU{} and N/A to IGN.
The companion Persistence classification restores applicability
at the N/A cells of Table~\ref{tab:master}.

The newer invariant-based construction instead derives one of
five verdicts:
N/A, BRK, MIX, PRS, or NEU.
It also distinguishes
a modeled N/A verdict from a matrix cell for which no derivation
has been supplied.

These representations are not interchangeable without an
explicit semantic correspondence.

$\PRS$ and $\NEU$ both designate non-breaking behavior in this
classification scheme, but they express different reasons.
A family-level value is meaningful only when its proposed behavior
is sufficiently uniform over the specified eligible operations;
otherwise the intended scope must be refined.
Neither the classification label nor its recorded presence in a
profile independently warrants strict identity for particular referents.

\begin{definition}[Non-breaking transformation]
  \label{se200.def.NonBreakingTransformation}
  A family $f$ is \emph{non-breaking} for profile $\tau$ when
  \[
    \Class_\tau(f)\in\{\PRS,\NEU\}.
  \]
  It is \emph{breaking} when $\Class_\tau(f)=\BRK$.
\end{definition}

\subsection{Identity Regimes and Generated Relations}
\label{subsec:identity_regimes}

An identity regime is a proposed pairing of an identity basis with a
transformation classification.
It is not itself an ontological kind
or proof that its basis is adequate for a particular referent.

\begin{definition}[Identity regime]
  \label{se200.def.IdentityRegime}
  An \emph{identity regime} $\tau$ consists of a declared identity basis
  and a classification function
  \[
    \Class_\tau:\Fset\to\{\PRS,\BRK,\NEU,\NA\}.
  \]
  The basis proposes conditions of sameness for the intended referent;
  the classification records proposed transformation behavior relative
  to that basis.
\end{definition}

\begin{definition}[Induced relation from non-breaking families]
  \label{se200.def.InducedIdentityRelation}
  Fix a profile $\tau$ and a domain $D_\tau$ of eligible presentations.
  Let $E_\tau\subseteq D_\tau\times D_\tau$ consist of pairs joined
  by specified, admissible steps of transformation families $f$
  satisfying
  \[
    \Class_\tau(f)\in\{\PRS,\NEU\}.
  \]
  Define $\sim_\tau$ as the smallest equivalence relation
  on $D_\tau$ containing $E_\tau$.
\end{definition}

This is a classification-based family-level construction.
It selects admissible steps according to the stipulated family
classifications; it does not independently establish that each
step preserves the declared identity basis.

The companion Persistence construction defines a different
family-level relation, generated only by steps of families
classified \PRS{}.
Under common domains and transformation dynamics, the
\PRS{}-only generated relation is contained in the relation
generated by \PRS{} and \NEU{} families.

The newer invariant-based construction instead evaluates
preservation at the level of individual admissible steps.
Its generating edges are exactly the steps that preserve
the declared invariant, including preserving steps of
families whose derived verdict is \MIX{}.
The three constructions must not be conflated.

Each generated relation is an equivalence relation by construction.
None independently establishes strict identity of the intended
real-world referents.
Such an interpretation requires an independently justified
identity criterion, evidence supporting the admissible steps,
and justification that equivalence closure preserves the criterion.

In particular, branching or a chain of classified non-breaking
steps cannot, by itself, discharge these scientific obligations.

\subsection{Basis Plurality}
\label{subsec:basis_plurality}

A carrier kind may admit more than one proposed identity basis.
For example, one can compare locus-fixed with object-fixed criteria,
coverage with structural organization, or rule content with rule structure.
These comparisons are well defined whichever basis an application adopts.

\begin{definition}[Basis plurality]
  \label{se200.def.BasisPlurality}
  A carrier kind exhibits \emph{basis plurality} within a construction
  when at least two distinct identity bases are specified for it.
  Whether both bases suit a given application is a separate question.
\end{definition}

The selected construction examines pairs of bases for plain referents
(locus or object), scopes (extension or structure), and rules
(content or structure).

\subsection{Hidden Regimes}
\label{subsec:hidden_regimes}

A representation can fail to disclose an identity-relevant distinction
on which its identity judgments nevertheless depend.
This issue is
separate from whether those judgments are expressed in one relation
or several relations.

\begin{definition}[Hidden regime]
  \label{se200.def.HiddenRegime}
  A \emph{hidden regime} is an identity-relevant criterion or distinction
  used to determine sameness in a realization but not adequately
  declared or fixed in the shared referential account, so that its
  application depends on an unstated or contested interpretation.
\end{definition}

A declared role, flag, carrier tag, or basis parameter is not hidden
merely because it is a discriminator.
Its adequacy depends on whether
its semantics are fixed and its use is justified for the intended
referents.
Conversely, an unacknowledged convention cannot be treated
as established common ground merely because a system implements it.
Whether a proposed combined representation is faithful must therefore
be examined rather than decided by terminology alone.

\subsection{Scope of the Analysis}
\label{subsec:scope_of_analysis}

The selected carrier inventory, identity bases, and transformation
classifications supply a finite candidate construction.
Sections~\ref{sec:algebra}--\ref{sec:distinctions} analyze its
transformation distinctions and the relations they generate.
The construction does not supply a complete inventory of identity
criteria or a general procedure for determining strict identity
under transformation.

The next section presents the selected reference inventory.
Its carrier
kinds identify subjects of inquiry; they do not themselves supply
independent identity warrants.


\section{Reference Inventory}
\label{sec:inventory}

Before analyzing identity under transformation, we select a finite
inventory of referents encountered in the accountability settings considered
in this paper.
These settings call for descriptions of parties, rules,
occurrences, scopes, records, and things acted upon or observed.
The inventory organizes questions of reference without adopting a
contested causal or normative interpretation as a substrate commitment.

The inventory is a working selection, not a theorem, an upper ontology,
or a complete domain ontology.
The kinds need not be mutually exclusive, and their names do not by
themselves supply criteria of individuation or persistence.

Reference kinds organize the intended referents; identity bases propose
conditions for their sameness under transformation.

\subsection{Reference Kinds and Identity Carriers}
\label{subsec:reference_kinds_identity_carriers}

\begin{definition}[Reference kind]
  \label{se200.def.ReferenceKind}
  A \emph{reference kind} is a class of intended referents selected to
  organize questions of reference in the accountability settings under
  analysis.
  Membership in a reference kind does not itself determine
  a referent's unity, sameness, or persistence criteria.
\end{definition}

\begin{definition}[Identity carrier]
  \label{se200.def.IdentityCarrier}
  An \emph{identity carrier} is an intended referent considered as the
  subject of a proposed sameness, difference, or persistence judgment
  under transformation.
  The designation does not itself warrant that
  judgment or determine the applicable identity basis.
\end{definition}

A reference kind can organize several proposed identity bases.
Different kinds may also support compatible criteria or share referents
considered in different capacities.
In either case, the adequacy of
an identity judgment depends on an identified bearer and an independently
justified criterion, not solely on a kind name or transformation label.

The selected inventory is:

\begin{center}
  \begin{tabular}{@{}l>{\raggedright\arraybackslash}p{0.68\textwidth}@{}}
    \toprule
    Reference kind & Referent considered in this analysis \\
    \midrule
    Obligation-bearer &
    a party considered as a possible bearer of responsibility, duty,
    liability, or authority \\
    Rule &
    a rule or normative instrument considered separately from an act
    of enacting, interpreting, or applying it \\
    Occurrence &
    a happening such as a filing, inspection, transfer, or decision,
    whose unity and identity require specified criteria \\
    Scope &
    a domain of applicability described by places, parties, periods,
    procedures, classifications, or conditions \\
    Record &
    a document, entry, or other informational record that can be
    referred to independently of what it describes or asserts \\
    Plain referent &
    a site, object, facility, instrument, sample, or asset considered
    apart from any assigned obligation-bearing or rule-governing role \\
    \bottomrule
  \end{tabular}
\end{center}

These categories draw on distinctions familiar in foundational ontology,
legal ontology, provenance, and institutional recordkeeping.
No novelty or independent necessity is claimed for this inventory.
Its role is to provide candidate carriers for the subsequent analysis.

\subsection{Obligation-Bearers}
\label{subsec:inventory_obligation_bearers}

The selected settings include claims about parties that may bear
responsibility, duty, liability, authority, or accountability.
The substrate can preserve reference to a party and record that a source
attributes an obligation to it without adopting that attribution as
an unqualified normative conclusion.

In multi-jurisdiction regulatory reporting, frameworks may identify
different entities or capacities as relevant: a parent organization,
subsidiary, facility, office, or delegated agent.
The organizational
entity, an institutional role, and the occupant of that role need not
have identical persistence criteria.
Whether two presentations concern
the same bearer requires an account appropriate to the intended referent.

Distinguishing a bearer from occurrences in which obligations are
assigned, contested, fulfilled, or discharged supports comparisons
across time.
One bearer may participate in several occurrences, and
one occurrence may involve several bearers.
These observations motivate
separate reference questions.

\subsection{Rules}
\label{subsec:inventory_rules}

The selected settings include reference to rules and normative
instruments that may be enacted, amended, delegated, interpreted,
applied, challenged, or repealed.
Recording such a referent does not
require the substrate to declare that the rule is valid, binding,
justified, or correctly interpreted.

Legislative history analysis distinguishes questions about a statute
from questions about the enactment, amendment, agency action, or
judicial decision associated with it.
A statute may remain citable
across those occurrences, but citation continuity alone does not
settle whether successive texts, versions, or normative contents
constitute the same rule.

The proposed content-fixed and structure-fixed bases examine different
responses to refinement.
Whether either basis supplies a warranted
identity criterion depends on which rule referent is intended and
what the transformation actually preserves.

\subsection{Occurrences}
\label{subsec:inventory_occurrences}

The selected settings include occurrences such as filings, inspections,
votes, transfers, transactions, observations, decisions, enactments,
amendments, and enforcement actions.
Their temporal realization and
provenance can help identify them but do not necessarily determine
where one occurrence begins or ends, which phases belong to one whole,
or when two reports concern the same happening.

For example, jurisdictions may disagree about whether a filing
constitutes compliance while referring to an occurrence they have
independently identified as the same filing.
Similarly, institutions
may dispute the significance of an enactment or a measurement without
thereby disputing its occurrence identity.
In other cases, that identity
itself may be contested and cannot simply be assumed as common ground.

An occurrence can be distinguished for analysis from its participants,
the records describing it, and the rules invoked in its evaluation.
Such analytic distinctions do not by themselves establish disjoint
ontological domains.

\subsection{Scopes}
\label{subsec:inventory_scopes}

The selected settings include applicability domains described by
places, parties, periods, procedures, classifications, or conditions.
Scopes may be territorial, institutional, temporal, procedural,
classificatory, jurisdictional, or conditional.
Their boundaries
and relevance may themselves be disputed.

In multi-jurisdiction reporting, a rule may be associated with
different entities, facilities, transactions, or reporting periods.
Tracking a scope separately in an analysis can permit comparison of
coverage while holding the rule description fixed, or comparison of
rule versions while holding coverage fixed.
This is a useful modeling
choice, not proof that every substrate must make scope an independent
identity carrier.

The proposed extension-fixed and structure-fixed bases distinguish
coverage from internal organization.
A decomposition may preserve
covered cases while changing organization; whether it preserves
identity depends on the justified scope criterion, not solely on
the operation's name.

\subsection{Records}
\label{subsec:inventory_records}

The selected settings include descriptive documents, dataset entries,
observation records, provenance items, and other informational records.
A substrate may record that a source asserted a claim without thereby
adopting its causal or normative content.

In cross-institutional provenance, a record may be produced, received,
annotated, copied, transferred, or reanalyzed.
The institutions may
disagree about what it shows.
Whether successive presentations are
of the same record nevertheless depends on the identity basis:
content, a particular token, a version, or a custody-linked object
may give different answers.

Distinguishing a record from what it describes permits separate
questions about an assertion and its subject.
It does not imply that
annotation, copying, or transfer always preserves record identity.

\subsection{Plain Referents}
\label{subsec:inventory_plain_referents}

The selected settings also include things acted upon, regulated,
located, transferred, observed, measured, or described: facilities,
parcels, instruments, devices, environmental sites, physical objects,
infrastructure components, samples, and assets.

The designation \emph{plain referent} indicates how a referent is being
considered in the present analysis, not a claim that the referent
could never bear an obligation, occupy a role, or appear in another
reference kind.
A facility may be inspected; a parcel may fall within
an applicability scope; an instrument may produce a measurement record;
and a sample may pass between institutions.
Whether the referent
persists through these operations remains a separate question.

The proposed locus-fixed and object-fixed bases distinguish questions
about a place from questions about an object associated with that
place.
A branch or fork may provide a useful test of their different
transformation responses, but does not by itself establish that they
are two identity criteria for one and the same kind of bearer.
The intended referents and eligible operations must be specified.

\subsection{Summary}
\label{subsec:inventory_summary}

The selected reference kinds organize a set of candidate identity
questions.
The distinction between a selected carrier and its proposed identity
basis remains essential: the inventory identifies intended referents,
and the bases propose criteria whose consequences can be tested
under transformation.

The next section compares proposed identity bases and transformation
behavior for these selected carriers under the formal setting of
Section~\ref{sec:formal_setting}.


\section{Transformation-Invariant Identity}
\label{sec:algebra}

The selected reference inventory of Section~\ref{sec:inventory} identifies
candidate carriers for comparing identity judgments under transformation.
This section examines how declared identity bases and classification profiles
can differ over the transformation basis of
Section~\ref{sec:formal_setting}.
A classification records proposed behavior relative to a specified basis;
it does not independently establish that two presentations have the same
real-world referent.
The scientific adequacy of the basis, the operation,
and their interpretation remains a separate obligation.

The inquiry draws on the distinction between reference kinds and identity
conditions familiar in formal ontology~\citep{guarino1998roles,guarinowelty2002}.
A kind name alone does not settle whether a transformation preserves its
intended bearer.
A stated identity basis helps formulate a test, but a
warranted persistence judgment also requires evidence and an appropriate
criterion for that referent.

\subsection{Identity Bases}
\label{subsec:identity_bases}

\begin{definition}[Identity basis]
  \label{se200.def.IdentityBasis}
  An \emph{identity basis} is a stated criterion proposed for deciding when
  eligible presentations concern the same intended referent.
  It specifies
  the conditions whose preservation would warrant sameness under the
  interpretation in question.
  Its declaration alone does not establish
  that the criterion is adequate for the intended referent.
\end{definition}

The selected construction considers single candidate bases for
obligation-bearers, occurrences, and records, and paired bases:
content versus structure for rules, extension versus structure for scopes,
and locus versus object for plain referents.
These are proposed criteria for examining transformation behavior.

A transformation classification is meaningful only relative to the
identity basis, the eligible comparison domain, and the actual effects
of the operation.
Different candidate bases may classify a selected
operation differently; that does not by itself decide which basis is
appropriate or whether the resulting judgments warrant strict identity.

\subsection{Profiles and Collapse}
\label{subsec:profiles_and_collapse}

Section~\ref{sec:formal_setting} defines identity regimes as proposed
basis--classification profiles and defines an equivalence relation
$\sim_\tau$ generated from specified non-breaking steps.
The generated relation and an independently warranted strict-identity
relation must not be conflated.

\begin{definition}[Collapse]
  \label{se200.def.Collapse}
  Fix a common comparison domain $D$ of eligible presentations for
  two profiles $\tau$ and $\tau'$.
  The profiles \emph{collapse at the level
    of generated relations on $D$} when the restrictions of their generated
  relations agree:
  \[
    {\sim_\tau}\big|_D = {\sim_{\tau'}}\big|_D.
  \]
  Those generated relations are \emph{distinct on $D$} if one relates
  an eligible pair that the other does not.
\end{definition}

Distinct classification functions do not necessarily induce distinct
generated relations.
For example, a family assigned \PRS{} under one profile and
\NEU{} under another is included under both profiles in
the classification-based \PRS{}-or-\NEU{} construction.
On common domains and transformation dynamics, its admissible
steps therefore contribute generating edges to both relations.

Those stipulated assignments do not independently establish
that every actual step preserves both declared invariants.

In the older family-generated construction, which uses only families
classified \PRS{}, distinct sets of preserving families need not yield
distinct relations on arbitrary transformation dynamics,
because the distinguishing families may have no relevant steps.
In the Persistence construction, profiles with identical preserving
sets generate identical relations on every shared dynamics
(Appendix~\ref{app:sep}).

A stronger result holds in the invariant-based step construction:
when an admissible step inside a common comparison domain preserves
one basis invariant but changes another, the two step-generated
relations differ.
The endpoints of the separating step cannot be connected by any path of
steps that each preserve the second basis, because equality of invariant
values is preserved along such a path
(Proposition~\ref{se200.prop.StepSeparation}).
This is a mathematical fact about the specified invariants and admissible steps,
not independent scientific warrant for strict identity of the intended referents.

\begin{definition}[Underdetermined kind]
  \label{se200.def.UnderdeterminedKind}
  A reference kind is \emph{underdetermined at a transformation family}
  $f$ within the selected construction when two proposed bases for that
  kind assign, on a shared eligible comparison domain, a non-breaking
  classification and a breaking classification, respectively, to $f$.
\end{definition}

This is a difference between candidate classification profiles.
If both bases are adequately justified for a particular use, a system must
retain the distinction between the judgments they authorize.
It does not follow that they require separately implemented or
non-amalgamable relations: an explicit basis parameter may preserve both.
Nor does a classification difference alone establish a separating pair
in the generated equivalence relations or in the world.

\subsection{Carrier Kinds Not Split in the Core Construction}
\label{subsec:core_unsplit_kinds}

The selected catalog specifies one candidate basis for each of
obligation-bearer, occurrence, and record.

For an \emph{obligation-bearer}, a possible criterion tracks the continuing
party rather than each obligation attributed to that party.
A change in location, assigned role, or obligation may preserve that party, provided
an appropriate account of the party's continuity warrants the judgment.
Replacement of the party may instead break its identity.
Organizational continuity and continuity of role occupancy must not be conflated.

For an \emph{occurrence}, temporal realization and provenance may supply
useful evidence, but they do not by themselves establish one bounded
whole.
Redescription of an independently individuated occurrence may
preserve it; treating a proper phase as the whole may not.
Decomposition, branching, and other transformations
cannot be classified uniformly without specifying
the occurrence and eligible presentations.

For a \emph{record}, content identity, token identity, version identity,
and continuity of custody can produce different judgments.
Copying a report may preserve its content while creating another token;
annotation may preserve an informational core while creating a new version.
Neither preservation nor breakage follows from the label ``record'' alone.

The selected profile names are
\[
  \OBL,\qquad \OCC,\qquad \REC.
\]
Their intended identity criteria and transformation classifications
remain subject to the qualifications above.

\subsection{Witnessed Split Kinds}
\label{subsec:witnessed_split_kinds}

The other carrier kinds in the selected catalog have pairs of
candidate bases.
Their comparisons can be expressed by a common
classification pattern.
In particular, a difference between proposed
values of $\Class_\tau(f)$ is an algebraic fact about the profiles;
a difference between warranted identity relations requires more.

\begin{lemma}[Split pattern for classification profiles]
  \label{se200.lem.SplitPattern}
  Let $K$ be a reference kind and let $\tau_1,\tau_2$ be two candidate
  profiles for $K$ over the same transformation basis and a shared
  eligible comparison domain.
  If some $f\in\Fset$ satisfies
  \[
    \Class_{\tau_1}(f)\in\{\PRS,\NEU\}
    \quad\text{and}\quad
    \Class_{\tau_2}(f)=\BRK,
  \]
  then the classification functions of $\tau_1$ and $\tau_2$
  are distinct.
  This conclusion concerns the classification functions; the relations
  they generate are compared in Section~\ref{sec:distinctions}.
\end{lemma}

\begin{proof}
  The first classification is different from $\BRK$, so the
  classification functions disagree at $f$ and cannot be equal.
  Their generated relations depend on the eligible step pairs and
  equivalence closure; those relations, and their interpretations,
  are not determined by this disagreement alone.
\end{proof}

A stronger, referent-level separation would need a common eligible
pair of presentations together with independently justified identity
criteria that agree and disagree on that pair.
The following examples
show how candidate separating situations can be constructed.
They do
not prove that the corresponding criteria are necessary for every
realization of the reference kind.

\subsection{Scope Under Decomposition}
\label{subsec:scope_split}

For a scope presentation, distinguish the cases covered from their
internal organization.
Consider a finite set $C=\{a,b\}$ and two
partitions
\[
  P=\{\{a,b\}\},\qquad P'=\{\{a\},\{b\}\}.
\]
The presentations $(C,P)$ and $(C,P')$ have the same extension but
different partition structures.
A stipulated extension-fixed basis
identifies them; a stipulated structure-fixed basis distinguishes them.
An eligible decomposition that replaces $P$ by $P'$ while preserving $C$
therefore provides a candidate separating example.

\begin{proposition}[Scope underdetermination in the selected model]
  \label{se200.prop.ScopeUnderdetermination}
  If an eligible $\mathrm{AD}$ operation preserves the coverage $C$
  while changing the identity-relevant partition $P$, then the
  extension-fixed and structure-fixed candidate profiles can assign
  different classifications to that operation.
\end{proposition}

\begin{proof}
  Under the stipulated extension-fixed criterion, equality of $C$
  supplies preservation; under the stipulated structure-fixed
  criterion, inequality of $P$ supplies breakage.
  On a domain where these are the specified effects of $\mathrm{AD}$,
  its two classifications differ, as in
  Lemma~\ref{se200.lem.SplitPattern}.
\end{proof}

We call the two proposed profiles
\[
  \SCOPEE,\qquad \SCOPES.
\]
This example demonstrates a distinction between coverage and
structure within the model.
It does not show that partition equality
is the correct identity criterion for every scope, or that a scope
cannot be faithfully represented with an explicit basis parameter.

\subsection{Rule Under Refinement}
\label{subsec:rule_split}

For a rule presentation, distinguish the specified normative content
from its textual or organizational structure.
Consider two
presentations $(N,S)$ and $(N,S')$ with identical stipulated content
$N$ but different structures $S\ne S'$.
A content-fixed comparison
identifies the presentations, whereas a structure-fixed comparison
separates them.
Whether a particular refinement actually preserves
normative content must be assessed rather than inferred from the word
``refinement.''

\begin{proposition}[Rule underdetermination in the selected model]
  \label{se200.prop.RuleUnderdetermination}
  If an eligible $\mathrm{RF}$ operation preserves the specified
  content $N$ while changing the identity-relevant structure $S$,
  then the content-fixed and structure-fixed candidate profiles can
  assign different classifications to that operation.
\end{proposition}

\begin{proof}
  The stipulated content-fixed criterion is preserved because $N$
  is unchanged; the stipulated structure-fixed criterion is broken
  because $S\ne S'$.
  The corresponding classifications differ,
  as in Lemma~\ref{se200.lem.SplitPattern}.
\end{proof}

We call the two proposed profiles
\[
  \RULEC,\qquad \RULES.
\]
This comparison does not decide whether a particular statute or rule
persists across amendment, recodification, or re-expression.
That
requires a justified account of the relevant rule referent.

\subsection{Plain Referent Under Forking}
\label{subsec:plain_referent_split}

For plain referents, distinguish a locus from an object associated
with that locus.
A toy configuration $(L,O)$ records both.
Two
configurations $(L,O_1)$ and $(L,O_2)$ may have the same locus $L$
but different objects $O_1\ne O_2$.
A locus-fixed comparison then
identifies them, while an object-fixed comparison distinguishes them.
The inequality of the objects is an assumption of the toy model,
not a conclusion from their labels alone.

\begin{proposition}[Plain-referent underdetermination in the selected model]
  \label{se200.prop.PlainReferentUnderdetermination}
  If an eligible $\mathrm{BF}$ operation yields compared presentations
  with the same locus but independently distinct objects, then
  locus-fixed and object-fixed candidate profiles can assign
  different classifications to that comparison.
\end{proposition}

\begin{proof}
  The locus-fixed criterion is preserved by the shared locus; the
  object-fixed criterion is broken by the stipulated object
  distinction.
  The candidate classifications consequently differ,
  as in Lemma~\ref{se200.lem.SplitPattern}.
\end{proof}

We call the two proposed profiles
\[
  \LOC,\qquad \OBJ.
\]
A fork of record representations does not by itself create distinct
objects or imply loss of object identity.
Nor does relocation of an
object necessarily end its identity.
The separating case depends on
which referent is being tracked and whether the assumed operation
and object distinction are independently warranted.

\subsection{Result of the Algebra}
\label{subsec:algebra_result}

The selected inventory contributes single-basis candidate profiles
for obligation-bearers, occurrences, and records, and paired profiles
for plain referents, scopes, and rules:
\[
  \begin{aligned}
     & \OBL,\quad \OCC,\quad \REC,                       \\
     & \LOC,\quad \OBJ,\quad \SCOPEE,\quad \SCOPES,\quad
    \RULEC,\quad \RULES.
  \end{aligned}
\]

The conditional examples show how different identity bases can
classify specified transformations differently.
They illustrate tests of the bases; they do not establish that every
transformation of a named family has the same effect.

Section~\ref{sec:profiles} collects the proposed classifications;
Section~\ref{sec:distinctions} compares the relations they generate and
shows how carrier-specific relations combine.


\section{Candidate Carrier--Basis Profiles}
\label{sec:profiles}

Sections~\ref{sec:inventory} and \ref{sec:algebra} yield a catalog of
named carrier--basis profiles, selected for analysis.

The following labels name the selected profiles:

\begin{description}[style=nextline,leftmargin=2.5cm]
  \item[$\OBL$] obligation-bearer identity.
  \item[$\OCC$] occurrence identity.
  \item[$\REC$] record identity.
  \item[$\LOC$] locus-fixed plain-referent identity.
  \item[$\OBJ$] object-fixed plain-referent identity.
  \item[$\SCOPEE$] extension-fixed scope identity.
  \item[$\SCOPES$] structure-fixed scope identity.
  \item[$\RULEC$] content-fixed rule identity.
  \item[$\RULES$] structure-fixed rule identity.
\end{description}

Table~\ref{tab:master} records proposed classifications over the finite
transformation basis $\Fset$.
Its entries are stipulated family-level profile assignments for the
selected comparison settings.
They must not be read as established
identity effects of every operation in the named family or as
independently warranted judgments about real-world referents.
Where eligible operations within a family have different effects,
that family requires a more specific classification domain.

\begin{table}[ht]
  \centering
  \small
  \setlength{\tabcolsep}{4pt}
  \begin{tabular}{@{}l ccc cc cc cc@{}}
    \toprule
       & OBL  & OCC  & REC  & LOC  & OBJ  & SCOPE-E & SCOPE-S & RULE-C & RULE-S \\
    \midrule
    RE & \NEU & \NEU & \NEU & \NEU & \NEU & \NEU    & \NEU    & \NEU   & \NEU   \\
    AN & \NEU & \NEU & \PRS & \NEU & \NEU & \NEU    & \NEU    & \NEU   & \NEU   \\
    RF & \PRS & \PRS & \PRS & \PRS & \PRS & \PRS    & \BRK    & \PRS   & \BRK   \\
    AD & \PRS & \BRK & \PRS & \PRS & \PRS & \PRS    & \BRK    & \PRS   & \BRK   \\
    RC & \NEU & \NEU & \NEU & \NEU & \NEU & \PRS    & \BRK    & \NEU   & \NEU   \\
    RA & \NEU & \NA  & \NEU & \NA  & \NA  & \NEU    & \BRK    & \NEU   & \NEU   \\
    SU & \BRK & \BRK & \BRK & \BRK & \BRK & \BRK    & \BRK    & \BRK   & \BRK   \\
    BF & \NEU & \NA  & \PRS & \PRS & \BRK & \PRS    & \BRK    & \BRK   & \BRK   \\
    PV & \NEU & \NEU & \PRS & \NEU & \NEU & \NEU    & \NEU    & \NEU   & \NEU   \\
    SE & \NEU & \PRS & \NEU & \PRS & \PRS & \NA     & \NA     & \NA    & \NA    \\
    \bottomrule
  \end{tabular}
  \caption{Proposed classification profiles over $\Fset$ for the
    selected carrier--basis construction.
    \PRS{}: preservation of the declared basis;
    \BRK{}: breakage of that basis;
    \NEU{}: applicable but identity-neutral under that basis;
    \NA{}: inapplicable on the selected domain.
    Entries are candidate assignments, not universal identity laws.}
  \label{tab:master}
\end{table}

A \PRS{} entry proposes preservation of the declared identity basis;
a \BRK{} entry proposes its breakage.
An \NEU{} entry indicates an applicable operation that does not affect
the basis, and an \NA{} entry indicates inapplicability on the selected
domain.
Neither \PRS{} nor \NEU{} independently warrants strict identity
between actual referents.
In particular, a \BRK{} entry in a proposed profile does not by itself
establish that two specified objects are distinct.

For the selected same-carrier comparisons, BF distinguishes the
proposed $\LOC$ and $\OBJ$ classifications; AD distinguishes
$\SCOPEE$ and $\SCOPES$; and RF distinguishes $\RULEC$
and $\RULES$.
The conditional examples in Section~\ref{sec:algebra} show how
stipulated locus, object, extension, structure, and content criteria
can yield these classification differences.
The other entries record additional proposed behavior whose adequacy
also requires examination; they are not needed to establish the
conditional split comparisons.

The uniform RE and SU rows are features of this selected table:
RE is assigned \NEU{} throughout, while SU is assigned \BRK{} throughout.
Those assignments are not general theorems about re-expression or
substitution under every scientifically adequate identity criterion.

A disagreement between two classification functions is distinct from
a disagreement between the equivalence relations generated by
non-breaking steps.
To state the latter, the two relations must be
compared on a common eligible domain.

\begin{definition}[Separation witness]
  \label{se200.def.SeparationWitness}
  Let $\tau$ and $\tau'$ be profiles with a common domain $D$ of
  eligible presentations.
  A \emph{separation witness for their generated
    relations on $D$} is a pair $(r,s)\in D\times D$ for which
  \[
    r\sim_\tau s \quad\text{and}\quad r\not\sim_{\tau'}s,
  \]
  or conversely, with each relation restricted to $D$.
\end{definition}

Such a witness demonstrates that the generated relations differ on
$D$; it does not independently establish the correctness of either
relation as strict real-world identity.
Nor does a difference in the
classification of a transformation family necessarily provide such
a witness, since equivalence closure may connect the relevant pair
by another non-breaking path.

\begin{proposition}[Distinct core profiles]
  \label{se200.prop.DistinctCoreProfiles}
  The nine classification functions specified by the columns of
  Table~\ref{tab:master} are pairwise distinct.
\end{proposition}

\begin{proof}
  It is enough to inspect the entries at AN, RF, RA, and BF.
  The four-entry classification signatures in these rows are
  pairwise different across all columns of
  Table~\ref{tab:master}.
  Hence no two of the
  classification functions agree on all of $\Fset$.
\end{proof}

This proposition concerns the \emph{declared classification functions}.
Whether their generated relations are pairwise distinct on a common
domain is taken up in Section~\ref{sec:distinctions} and
Appendix~\ref{app:sep}.

\subsection{Initial Model-Derived Classification Results}
\label{subsec:model_derived_results}

Table~\ref{tab:master} records the proposed classification matrix.
Its entries are not inputs to the following derivations.

Instead, the companion Lean models~\citep{case2026identityregimes}
specify presentation fields, identity invariants,
admissible transformation steps, and their effects.

An invariant basis consists of a function on presentations
and a declared support of presentation fields.
Agreement on the supported fields implies equality of the invariant.
The Lean development additionally proves minimality of the
supports used in the certified cells: every supported field
can independently affect its invariant.

An effect model declares admissible transformation steps and
the fields each transformation family may write.
An effective instance is an admissible step that changes
at least one presentation field.

\begin{definition}[Model-derived transformation verdict]
  \label{se200.def.ModelDerivedVerdict}
  Fix an invariant basis $b$, an effect model $M$, and a
  transformation family $f$.

  The derived verdict is:
  \begin{enumerate}[label=(\alph*),nosep]
    \item $\NA$: no effective instance exists;
    \item $\BRK$: effective instances exist and every one
          changes the invariant;
    \item $\MIX$: some effective instance changes the invariant
          and another preserves it;
    \item $\PRS$: effective instances exist, all preserve the
          invariant, and at least one changes a supported field;
    \item $\NEU$: effective instances exist, all preserve the
          invariant, and none changes a supported field.
  \end{enumerate}
\end{definition}

\MIX{} establishes that the declared transformation family contains
both invariant-preserving and invariant-breaking effective instances.

The family therefore cannot receive a uniform preserving or breaking
verdict under those modeled conditions.
A uniform verdict would require restricting or refining the family.

Alternatively, individual preservation judgments can be retained
at the step level without assigning one uniform verdict to the family.

Applicability is defined for the model's presentation space,
not for an arbitrarily selected smaller comparison domain.

\begin{table}[ht]
  \centering
  \small
  \setlength{\tabcolsep}{5pt}
  \begin{tabular}{@{}p{0.30\linewidth}p{0.41\linewidth}cc@{}}
    \toprule
    Pair and family       & Model or explicit warrant                                             & First & Second \\
    \midrule
    $\SCOPEE/\SCOPES$, AD & Coverage-preserving; one organization field                           & \PRS  & \BRK   \\
    $\SCOPEE/\SCOPES$, AD & Coverage-preserving; field per case-pair                              & \NEU  & \BRK   \\
    $\SCOPEE/\SCOPES$, AD & Broader variant allowing coverage changes; field per pair             & \MIX  & \BRK   \\
    $\RULEC/\RULES$, RF   & Norm-preserving strict refinement; coarse organization field          & \PRS  & \BRK   \\
    $\RULEC/\RULES$, RF   & Norm-preserving strict refinement; fields per case-pair and case norm & \NEU  & \BRK   \\
    $\RULEC/\RULES$, RF   & Broader variant permitting changes to norms                           & \MIX  & \BRK   \\
    $\LOC/\OBJ$, BF       & Branch changes; no further warrant                                    & \MIX  & \MIX   \\
    $\LOC/\OBJ$, BF       & Colocated and FreshObject warrants                                    & \NEU  & \BRK   \\
    $\LOC/\OBJ$, BF       & Colocated and SameObject warrants                                     & \NEU  & \NEU   \\
    \bottomrule
  \end{tabular}
  \caption{Selected verdicts under explicitly specified Lean effect
    models and warrants.
    The two rightmost columns follow the order of the bases named
    in the first column.
    The field-level strict AD, strict RF, and warranted BF rows
    account for the certified cells in the derived matrix.
    The remaining rows are supplementary model comparisons.
    All verdicts are conditional mathematical results,
    not universal identity laws.}
  \label{tab:derived_initial}
\end{table}

The certified cells in the derived matrix of the companion Lean
development are:

\begin{itemize}
  \item $\SCOPEE$ under strict AD: \NEU;
  \item $\SCOPES$ under strict AD: \BRK;
  \item $\RULEC$ under strict RF: \NEU;
  \item $\RULES$ under strict RF: \BRK;
  \item $\LOC$ under BF with Colocated and FreshObject warrants: \NEU;
  \item $\OBJ$ under BF with those same warrants: \BRK.
\end{itemize}

Each certified cell contains an explicit effect model,
an invariant basis with a proven minimal support,
and a proof of its derived verdict.
The remaining cells in the stipulated matrix
have no derived model yet.
For every certified cell, the derived and stipulated
classifications agree on whether the transformation is breaking.

The derived \NEU{} cells are classified \PRS{}
in Table~\ref{tab:master}.
Both verdicts are non-breaking, but they distinguish
whether a transformation engages a supported field.

The scope and rule comparisons expose dependencies on
field granularity.
The fork classification additionally depends on explicit
continuity warrants.
These conditional results do not establish a universal
correspondence between stipulated and derived classifications.

The additional rows in Table~\ref{tab:derived_initial}
are sensitivity analyses or alternative warranted models,
not additional certified matrix cells.

Two findings are especially consequential.
First, refining the scope presentation from one organization field
to one field per case-pair changes the strict AD extension verdict
from \PRS{} to \NEU{}.
The fine-grained extension basis reads diagonal entries (coverage),
while restructuring changes only off-diagonal entries.
The support of each fine-grained scope basis is proved minimal in that
model.
Both verdicts are non-breaking, but the classification is
not invariant under the choice of field representation.
The model therefore exposes a granularity dependence that must be
controlled before an unqualified classification can be asserted.

The rule model exhibits a corresponding dependence.
Under strict RF, the content basis is \PRS{} when the organization
is represented coarsely, but \NEU{} when fields separately
represent case-pair organization and case-specific norms.
The structure basis is \BRK{} in both representations.
The field-level content support reads diagonal organization
entries and the case-specific norms; strict RF preserves both.
The content and structure supports are proved minimal in this model.

Second, branching alone specifies a change of branch, not a change of
locus or object.
With no additional warrant, both corresponding verdicts are \MIX.
Under the Colocated and FreshObject warrants,
the locus basis is neutral and the object basis breaking;
under the Colocated and SameObject warrants,
both are neutral and their step-generated relations
collapse on every comparison domain.
These outcomes are consequences of distinct, explicitly stated
hypotheses.
None establishes that one warrant is true of every fork
in an actual system.

The broader scope variant in the table permits changes to coverage and
therefore is not an instance of the coverage-preserving AD family as
originally defined in Section~\ref{sec:formal_setting}.
It is included as a separate sensitivity analysis.
The broader rule variant likewise changes the specification of eligible refinements.
A comparison with Table~\ref{tab:master} is warranted only after
matching operation semantics, identity basis, field granularity, and
admissibility conditions.

The invariant-based step construction uses a presentation space $C$,
an effect model $M$, an invariant basis $b$, and a comparison domain
$D\subseteq C$.

Define the preserving-step relation by
\[
  R_{M,b,D}(x,y)
  \iff
  x,y\in D
  \;\land\;
  \exists f\,
  \bigl(
  M.\mathrm{step}(f,x,y)
  \land
  b.\mathrm{inv}(x)=b.\mathrm{inv}(y)
  \bigr).
\]

The step-generated identity relation is
\[
  \mathrm{StepId}_{M,b,D}
  =
  \operatorname{EqCl}_{C}(R_{M,b,D}),
\]
where $\operatorname{EqCl}_{C}$ is the smallest equivalence
relation on $C$ containing the specified generating steps.

Unlike the classification-based family-level constructions,
this relation tests invariant preservation for each admissible
step. Its generating edges may include preserving steps from
families whose derived verdict is \MIX{}.

\begin{theorem}[Step-generated relation equality]
  \label{se200.thm.StepIdentityCollapse}
  Fix a common effect model, two invariant bases $b$ and $b'$,
  and a comparison domain $D$.

  The equivalence relations generated by their respective
  invariant-preserving admissible steps are equal if and only if
  the bases agree on invariant preservation for every admissible
  step with both endpoints in $D$.
\end{theorem}

\begin{proof}
  If preservation judgments agree on every admissible step
  in $D$, the generating relations are equal.
  Their equivalence closures therefore agree.

  Conversely, suppose the generated relations agree.
  An admissible step preserving $b$ belongs to the relation
  generated by $b$ and hence also to the relation generated
  by $b'$.

  Every generating step for the latter relation preserves
  $b'$. Equality of invariant values is an equivalence
  relation, so its closure also preserves $b'$.
  Thus the original step preserves $b'$.
  The converse follows symmetrically.
\end{proof}

\begin{proposition}[Separating transformation step]
  \label{se200.prop.StepSeparation}
  Let $x,y\in D$ be connected by an admissible
  transformation step.

  If $b(x)=b(y)$ but $b'(x)\ne b'(y)$, then the
  step-generated identity relations for $b$ and
  $b'$ are distinct.
\end{proposition}

\begin{proof}
  The step generates a relation between $x$ and $y$
  under $b$.

  Under $b'$, the generated relation cannot connect
  $x$ and $y$, because every generating step preserves
  $b'$, and equality of invariant values is preserved
  under equivalence closure.
\end{proof}

These derivations are mathematically conditional.

Independently justifying the invariants and fork warrants for actual referents,
and showing that a generated equivalence relation expresses warranted
strict identity, remain separate scientific obligations.


\section{Profile Distinctions and Carrier-Indexed Combination}
\label{sec:distinctions}

The preceding sections specify candidate carrier kinds,
identity bases, and transformation classifications.
This section compares the resulting profiles by their classification
functions and by the relations they generate, and shows how
carrier-specific relations can be combined in one tagged relation.

\subsection{Witnessed Differences Within the Selected Profiles}
\label{subsec:witnessed_splits}

The proposed core contains pairs of identity bases associated
with the same carrier kind.
These pairs motivate tests of whether ordinary transformations make
different identity judgments under the respective bases.

For plain referents, the proposed locus-fixed and object-fixed bases
are compared using branch/fork operations
(Proposition~\ref{se200.prop.PlainReferentUnderdetermination}).
A particular fork may preserve a locus while distinguishing the objects
associated with that locus.
The question for a realization is which bearer each basis tracks and
whether the proposed preservation and breakage judgments are warranted.

For scopes, the proposed extension-fixed and structure-fixed bases
are compared using aggregation and decomposition
(Proposition~\ref{se200.prop.ScopeUnderdetermination}).
An operation may preserve coverage while changing component structure.
Whether that difference is an identity difference depends on the
independently justified criterion for the scope in question.

For rules, the proposed content-fixed and structure-fixed bases
are compared using refinement
(Proposition~\ref{se200.prop.RuleUnderdetermination}).
A refinement may preserve specified content while changing structure.
Again, the relevant sameness judgment depends on what the rule is
and on which criterion of its identity is justified.

Two justified identity relations that disagree on an eligible pair
of presentations are distinct on that comparison domain.
A difference between declared transformation classifications does not
by itself exhibit a pair on which the generated relations disagree.
Such a witness requires a shared eligible comparison domain and
appropriate transformation and identity interpretations.
These pairs supply candidate distinctions for further testing.

\subsection{Profiles and Generated Relations}
\label{subsec:profiles_vs_relations}

The profiles catalogued in Section~\ref{sec:profiles}
arise from the selected carrier--basis assignments.
Proposition~\ref{se200.prop.DistinctCoreProfiles} concerns distinctions
among their classification functions.

The number of distinct family-level generated relations
depends on the generating rule.
The \PRS{}-only and \PRS{}-or-\NEU{} constructions
induce different patterns of equal family sets;
Appendix~\ref{app:sep} enumerates them and gives the resulting
upper bounds.
Further equalities may arise when the dynamics do not distinguish
the selected family sets.
These bounds do not apply to the invariant-based step relations,
whose equality and separation are characterized by
Theorem~\ref{se200.thm.StepIdentityCollapse} and
Proposition~\ref{se200.prop.StepSeparation}.

\subsection{Carrier-Indexed Combination}
\label{subsec:amalgamation}

An elementary construction shows that relations on distinct carrier
kinds can be represented together while preserving each component
equivalence relation.
Let $K$ index the selected carrier kinds, and let $D_k$ be a domain
of presentations for carrier kind $k$.
Suppose $E_k$ is an equivalence relation on $D_k$ for each $k\in K$.
Use the tagged disjoint union
\[
D = \bigsqcup_{k\in K} D_k
\]
and define a relation $E$ on $D$ by
\[
(k,x)\mathrel{E}(\ell,y)
\quad\Longleftrightarrow\quad
\bigl(k=\ell \;\wedge\; x\mathrel{E_k} y\bigr).
\]

\begin{proposition}[Carrier-indexed combination]
\label{se200.prop.CarrierIndexedCombination}
Then $E$ is an equivalence relation on $D$.
Restricting $E$ to the $k$-th summand recovers $E_k$
for each $k\in K$.
\end{proposition}

\begin{proof}
Reflexivity and symmetry follow from the corresponding properties
of each $E_k$.
If $(k,x)\mathrel{E}(\ell,y)$ and
$(\ell,y)\mathrel{E}(m,z)$, then $k=\ell=m$ and
$x\mathrel{E_k}y\mathrel{E_k}z$.
Transitivity of $E_k$ therefore gives $(k,x)\mathrel{E}(m,z)$.
For presentations in the $k$-th summand, the definition of $E$
is exactly the definition of $E_k$.
\end{proof}

This construction preserves the component equivalence relations
without identifying presentations belonging to different carrier kinds.
Distinct carrier kinds therefore do not by themselves require
separate formal structures: a declared carrier tag carries the
distinction within one relation.
Whether a combined relation is an adequate identity account, preserves
all required transformation semantics, and satisfies the Neutral
Substrate constraint depends on the semantics of the tags and on the
bases proposed for each carrier kind.
A publicly declared carrier or basis parameter is not a hidden regime
merely because it is a parameter; its adequacy depends on whether its
semantics are fixed and its use is justified.

\subsection{Separation Under an Extended Transformation Basis}
\label{subsec:monotonicity}

A narrower monotonicity result applies to the declared
classification profiles.
Let $\Fset\subseteq\Fset'$ be transformation bases.
Suppose the classifications of two profiles on $\Fset'$ agree
with their existing classifications on every operation in $\Fset$.

\begin{proposition}[Persistence of a classification distinction]
\label{se200.prop.ClassificationDistinctionPersists}
If two profiles have different classifications for some
$f\in\Fset$, their classification functions remain distinct
under extensions to $\Fset'$ that preserve the original classifications.
\end{proposition}

\begin{proof}
The extended classification functions still disagree at $f$,
so they cannot be equal.
\end{proof}

This result records that a classification difference, once established,
is stable under extensions of the transformation basis that retain the
original classifications.

The broader limitations are discussed in Section~\ref{sec:limits}.


\section{Determinacy of the Selected Profile Assignment}
\label{sec:determinacy}

Sections~\ref{sec:inventory}--\ref{sec:profiles} specify a finite catalog of
carrier--basis pairs and proposed transformation-classification profiles.
This section asks whether the catalog assigns a named profile to each
selected pair without ambiguity.
This is an internal property of the specified assignment.

\subsection{Core Carrier--Basis Pairs}
\label{subsec:core_carrier_basis_pairs}

\begin{definition}[Core carrier--basis pair]
\label{se200.def.CoreCarrierBasisPair}
A \emph{core carrier--basis pair} is one of the selected carrier-kind
and candidate identity-basis combinations displayed below.
The arrows
record the profile names assigned within this construction:
\[
\begin{array}{rcl}
\text{obligation-bearer with responsibility-bearing identity} &\mapsto& \OBL,\\
\text{occurrence with realization-and-provenance identity} &\mapsto& \OCC,\\
\text{record with descriptive-record identity} &\mapsto& \REC,\\
\text{plain referent with locus-fixed identity} &\mapsto& \LOC,\\
\text{plain referent with object-fixed identity} &\mapsto& \OBJ,\\
\text{scope with extension-fixed identity} &\mapsto& \SCOPEE,\\
\text{scope with structure-fixed identity} &\mapsto& \SCOPES,\\
\text{rule with content-fixed identity} &\mapsto& \RULEC,\\
\text{rule with structure-fixed identity} &\mapsto& \RULES.
\end{array}
\]
\end{definition}

The carrier kinds listed first have one candidate basis each;
the others have two, motivated by the conditional comparisons in
Section~\ref{sec:algebra}.

\begin{definition}[Core regime assignment]
\label{se200.def.CoreRegimeAssignment}
The \emph{core regime assignment} $A$ maps each selected core
carrier--basis pair to the named classification profile indicated by
the corresponding arrow in Definition~\ref{se200.def.CoreCarrierBasisPair}.
Its codomain is the selected profile catalog in Section~\ref{sec:profiles}.
\end{definition}

The assignment is stipulated as part of the finite construction.
It does not choose or justify a basis for an actual referent.

\subsection{Existence}
\label{subsec:determinacy_existence}

\begin{lemma}[Assignment existence]
\label{se200.lem.CoreExistence}
Every selected core carrier--basis pair has an assigned profile under $A$.
\end{lemma}

\begin{proof}
Definition~\ref{se200.def.CoreCarrierBasisPair} lists all selected
pairs and gives an image for each.
By
Definition~\ref{se200.def.CoreRegimeAssignment}, $A$ is defined on
every pair in its stated domain.
\end{proof}

\subsection{Uniqueness}
\label{subsec:determinacy_uniqueness}

\begin{lemma}[Assignment uniqueness]
\label{se200.lem.CoreUniqueness}
Each selected core carrier--basis pair has exactly one assigned profile
under $A$.
\end{lemma}

\begin{proof}
Each pair occurs exactly once on the left-hand side of the displayed
assignment in Definition~\ref{se200.def.CoreCarrierBasisPair}, with
one profile name on the right-hand side.
The definition of $A$ assigns
no additional value to that pair.
Hence its image under $A$ is unique.
\end{proof}

\subsection{Determinacy Theorem}
\label{subsec:determinacy_theorem}

\begin{theorem}[Core regime assignment]
\label{se200.thm.CoreRegimeAssignment}
For each carrier--basis pair in the selected catalog, there exists
exactly one named classification profile assigned to it by $A$.
\end{theorem}

\begin{proof}
Existence is Lemma~\ref{se200.lem.CoreExistence}; uniqueness is
Lemma~\ref{se200.lem.CoreUniqueness}.
\end{proof}

This theorem records determinacy of a defined, finite assignment.
It does not derive the assignment from the neutrality constraint,
establish the scientific correctness of its classification entries,
or show that the assigned profile yields a warranted strict-identity
relation on real-world referents.
Other candidate profiles or identity
bases may be relevant outside this selected catalog.

\subsection{Limits of Basis Declaration}
\label{subsec:basis_declaration_failure}

A carrier-kind label alone does not select one profile when the
catalog contains more than one candidate basis for that kind.
For example, a scope may be examined under extension-fixed or
structure-fixed criteria, a rule under content-fixed or
structure-fixed criteria, and a plain referent under locus-fixed
or object-fixed criteria.
Supplying the corresponding basis label
makes the input to $A$ unambiguous \emph{within this catalog}.

That is not the same as establishing which basis is appropriate for
a particular referent.
A declared basis needs independently justified
individuation and persistence conditions, and each proposed
transformation classification needs an adequate interpretation.
Even the carrier kinds represented by only one profile in this
catalog may admit other legitimate identity bases.

Consequently, a possible conformance check could verify whether a
system discloses a selected basis and applies its declared profile
consistently.
It could not, on that evidence alone, certify the
basis's scientific adequacy or the truth of its identity judgments.
This is the boundary of the determinacy result proved here.


\section{Applications as Identity-Criterion Tests}
\label{sec:application}

Sections~\ref{sec:inventory}--\ref{sec:determinacy} define a selected
inventory, candidate carrier--basis profiles, and their stipulated
transformation classifications.
The following settings illustrate questions
that those distinctions can help formulate.

For each setting, the relevant questions are which referent is intended,
which presentations are eligible for comparison, what changes under a
transformation, and what evidence justifies sameness or difference.
Where a common referential judgment cannot be warranted, the Neutral
Substrate construction does not license adopting it as an unqualified
foundational commitment merely because the records share a label.

\subsection{Multi-Jurisdiction Regulatory Reporting}
\label{subsec:app_regulatory}

A regulatory record may refer to a corporation, a subsidiary, an office,
or a party occupying an office.
The selected $\OBL$ profile offers a
candidate basis for following an obligation-bearer across changes in
reported duties and activities.
Whether the same legal or institutional
bearer persists requires an appropriate continuity criterion; sharing a
name, role, or identifier is not sufficient.
Different jurisdictions may
also identify different bearers for a purported obligation.
The shared
record must distinguish an identified party from a framework's attribution
of responsibility to that party.

A reporting statute is a candidate rule referent.
Under $\RULEC$, the
comparison asks whether specified normative content is preserved; under
$\RULES$, it asks whether a specified textual or internal structure is
preserved.
Re-codification might preserve requirements while changing
section organization, yielding different \emph{proposed} classifications.
That comparison does not decide whether the resulting enactment is the
same statute as a legal object.
Its legal continuity, the meaning of its
requirements, and the intended unit of reference must be justified
independently.

Filings, inspections, transactions, and enforcement actions provide
candidate occurrences for $\OCC$.
A filing occurrence is distinguishable
from the filed document and from subsequent copies or versions of that
document, which can be examined under $\REC$.
Two jurisdictions may agree
which filing occurred while disagreeing whether it constituted compliance.
They may also disagree about whether two reported activities are the same
occurrence.
In the latter case, common occurrence identity cannot simply
be assumed from matching timestamps or document identifiers.

A jurisdictional scope can be examined by the cases it covers
($\SCOPEE$) or by a specified administrative or procedural structure
($\SCOPES$).
Dividing a reporting domain into districts may preserve
its coverage while changing its partition.
This supplies a conditional
transformation comparison, not a general rule that the jurisdiction
itself persists under subdivision.
The identity of a legal jurisdiction
may depend on further institutional facts not captured by either profile.

A regulated facility, parcel, or instrument supplies a possible plain
referent.
The $\LOC$ comparison follows a site or location, whereas
$\OBJ$ follows the object occupying or moving through it.
A facility
relocated from one site to another need not cease to be the same
facility; conversely, continued use of one site does not make successive
occupants one object.
The proposed basis must match the intended bearer.

These comparisons make all selected carrier kinds available for
analysis and illustrate the candidate split pairs.
A particular system may need only some of the candidate profiles.
Agreement about a carrier's designation does not by itself supply the
identity criterion needed to compare it across jurisdictions.

\subsection{Legislative History Tracking}
\label{subsec:app_legislative}

A statute can be tracked by its recorded requirements, its enacted text,
its section structure, or the legal continuity of the instrument across
amendments.
The selected $\RULEC$ and $\RULES$ profiles isolate content-
and structure-based comparisons.
An amendment may retain specified
requirements while reorganizing sections; a textual continuation may
also be assigned a changed interpretation.
Neither observation alone
establishes the strict identity of the statute before and after the
amendment.
The referent and its identity conditions must first be fixed.

Enactments, votes, amendments, applications, and judicial decisions are
candidate occurrences for $\OCC$.
An enactment is not the enacted rule,
and an account of an enactment is not the enactment itself.
Legislative
texts, committee reports, docket entries, and interpretive memoranda
supply record candidates for $\REC$.
Copies with identical content may
be distinct documentary tokens, and later versions may differ without
settling whether they remain the same informational object.
The record
basis must therefore be specified rather than presumed.

Legislatures, agencies, courts, officers, and regulated parties can be
considered as obligation- or authority-bearing parties under $\OBL$,
provided the intended bearer is distinguished from an office, its
occupant, or a particular institutional role.
The same office may have
successive occupants; the same person may move to another office.
Such changes do not, by themselves, settle the identity of either the
person or the office.

Geographic, procedural, temporal, and institutional boundaries provide
candidate scopes.
A redistricting operation can preserve a chosen set
of covered cases while changing a partition, offering a comparison
between $\SCOPEE$ and $\SCOPES$.
Objects affected by the legal regime,
such as land parcels, facilities, or instruments, may be considered
under $\LOC$ or $\OBJ$ when those bases are appropriate.

Legislative history therefore offers tests for whether distinctions
among an instrument, its events, its documents, its institutional
bearers, and its scopes can be maintained without adopting contested
normative conclusions.
The formal profile catalog organizes these
tests; it does not decide which legal theory of continuity is correct.

\subsection{Cross-Institutional Provenance}
\label{subsec:app_provenance}

Cross-institutional provenance highlights the distinction between a
record and what the record purports to describe.
A dataset entry,
measurement report, transfer log, annotation, or derived metric is a
candidate for $\REC$.
A copy may preserve content while creating a
new token; an annotation or derivation may instead create a new version
or informational object.
Neither common content nor an unbroken
provenance chain alone settles which record-identity criterion applies.

A sampled object, instrument, facility, or collection site is a
candidate plain referent.
The $\LOC$ profile asks about continuity of
a specified site or position; the $\OBJ$ profile asks about continuity
of a specified object.
Forking record representations does not
necessarily fork the represented object.
Conversely, a process that
separates, replaces, consumes, or transforms physical material may
raise identity questions that preservation of a location or inventory
label cannot resolve.
The intended referent and transformation effects
must be established separately.

Producing organizations, custodians, laboratories, agencies, and
recipients supply possible $\OBL$ carriers where their continuity and
responsibility-bearing status are independently specified.
Transfers,
measurements, analyses, approvals, and custody changes supply possible
$\OCC$ carriers.
A transfer log is a record of a proposed transfer
occurrence, not automatically that occurrence or proof of its identity.

Data-use policies, consent terms, retention rules, and access
restrictions provide candidate rule referents for $\RULEC$ or
$\RULES$.
Institutional coverage conditions, consent domains, and
data-sharing boundaries provide candidate scopes for $\SCOPEE$ or
$\SCOPES$.
Whether changes to policy text, permitted uses, scope
coverage, or organizational structure preserve the relevant referent
remains dependent on justified identity criteria.

These distinctions help separate shared reference from disagreements
about data quality, classification, causation, responsibility, or
permissible use.
They also expose circumstances in which institutions
lack enough warranted common reference for the Neutral Substrate
assumption to be discharged.
In that case, disputed co-reference or
persistence judgments should not be treated as established merely
because the records can be joined.

\subsection{Summary}
\label{subsec:app_summary}

The settings illustrate the analytic use of the selected
carrier--basis catalog: $\OBL$, $\OCC$, and $\REC$ organize questions
about bearers, occurrences, and records; $\LOC/\OBJ$,
$\SCOPEE/\SCOPES$, and $\RULEC/\RULES$ organize candidate comparisons
between different bases for plain referents, scopes, and rules.
Their usefulness lies in keeping these questions explicit.
The examples do not independently validate the table's transformation
classifications.
To assess scientific adequacy, a concrete study must
identify the intended referents and eligible presentations, justify
its proposed identity criteria, and evaluate the resulting judgments
against relevant examples and counterexamples.

Reference kinds select questions for analysis; declared bases propose
criteria; transformation profiles state candidate consequences.
Strict identity and persistence under transformation require further
warrant, and the conditions for combining such criteria without loss
remain open.


\section{Related Work}
\label{sec:related}

This paper is situated at the intersection of formal ontology, provenance,
and classification under disagreement.
It does not propose a new upper ontology.
It compares selected carrier--basis profiles under specified
transformations and derives how their classifications and generated
relations behave.

\subsection{Classification under Disagreement}
\label{subsec:related_classification_disagreement}

Classification systems can embed institutional, social, and evaluative
commitments rather than functioning as neutral descriptions~\citep{bowkerstar1999,haslanger2012}.
In scientific and policy settings, disagreement may persist because
participants adopt different evidentiary, methodological, or evaluative
frameworks, not merely because a temporary empirical uncertainty remains~\citep{longino1990}.
These observations motivate shared representations that do not presuppose
interpretive convergence.

The present paper takes this motivation from \emph{Neutral Substrates}~\citep{case2026neutral}.
That work establishes a conditional sufficiency result for semantic
neutrality under Referential Common Ground; it does not supply
independently justified identity criteria for every intended referent.
Here the narrower question is how proposed identity bases respond to
specified record transformations, and what evidence would justify their
use for shared reference.

\subsection{Identity, Roles, and Ontological Methodology}
\label{subsec:related_identity_roles}

Formal ontology distinguishes identity criteria from ordinary
classification and emphasizes the need to examine how roles and other
properties relate to the identity of their bearers.
OntoClean, in particular, analyzes rigidity and identity-related
constraints on taxonomic organization~\citep{guarino1998roles,guarinowelty2002}.
Accounts of social and institutional entities further examine bearers,
roles, social objects, and institutional structures~\citep{masolo2003wonderweb,guizzardi2005}.

This paper uses those distinctions as motivation, not as a derivation
of its proposed identity bases.
A role can be relevant to representing
or tracking a referent without independently supplying a sufficient
criterion for that referent's strict identity.
Conversely, the use of
an explicit role or basis parameter does not by itself make a
representation inadequate.
The question is which referent is intended,
what conditions justify treating presentations as the same referent,
and what a transformation actually preserves.

The paper compares declared transformation classifications under
candidate identity bases.
A difference between classifications is a
formal distinction, but it does not alone prove a difference between
generated equivalence relations or scientifically warranted identities.

\subsection{Upper Ontologies}
\label{subsec:related_upper_ontologies}

Upper ontologies such as BFO, DOLCE, and UFO provide resources for
modeling entities, occurrences, dependence, qualities, roles, and
institutional or social objects~\citep{smithceusters2015bfo,arp2015building,gangemi2002dolce,guizzardi2005}.
They address broader ontology-engineering purposes than the finite
comparison undertaken here.
The profiles are not offered as additions to those ontologies, which
may already make the distinctions.

For example, a BFO-based model might represent a particular occurrence
using an occurrent category and a material object using a continuant
category, subject to the ontology's actual modeling commitments.
Rules, records, and institutional bearers likewise require analysis of
the intended referents and the resources available in a chosen ontology.
Such mappings cannot be assumed solely from names in the selected
inventory.
They must be justified in the application concerned.

DOLCE- and UFO-based representations similarly offer resources for
modeling social objects, roles, qualities, and dependence relations.
A realization built with any of these frameworks can be examined for
whether it distinguishes, where appropriate, a rule's content from its
structure, a scope's coverage from its organization, a location from
an occupying object, and a record from the thing recorded.
Whether
those are identity distinctions, rather than differences of description
or representation, remains dependent on the selected referent and
independently justified identity criterion.

Sections~\ref{sec:algebra}--\ref{sec:distinctions} compare such candidate
distinctions.
The comparison can help make modeling assumptions explicit and expose
cases needing further scientific analysis.

\subsection{Provenance, Versioning, and Exchange}
\label{subsec:related_provenance_exchange}

The transformation families used here resemble operations encountered
in provenance and versioning systems.
PROV provides vocabulary for
activities, entities, derivation, attribution, generation, use, and
revision~\citep{provdm2013}.
These relations may supply evidence about
how representations were produced or changed.
They do not, without
additional identity conditions, settle whether two records are the
same record, copies of a record, or records about the same referent.

An exchange formalism such as Common Logic may provide a setting for
expressing some of the relevant commitments~\citep{isoiec24707}.
Conservative extension and heterogeneous ontology exchange are
related logical concerns~\citep{hodges1997,kutz2010}.
However,
compatibility of a substrate with admissible frameworks is not, by
itself, a proof of conservativity, and this paper constructs neither
a Common Logic encoding nor a conservative-extension theorem.
Its formal results concern the selected transformation profiles and
the limits of conclusions drawn from them.

Related questions about long-lived reference and reuse arise in
information infrastructures and digital-twin systems~\citep{edwards2011,grieves2016}.
The candidate profiles offer questions to test in such settings:
what is the intended bearer, which transformations preserve it,
and what available evidence warrants those judgments?
The paper does not establish a generally applicable solution for
those systems.

\subsection{Causal and Normative Separation}
\label{subsec:related_causal_normative}

The distinction between observations or records and causal claims is
important in causal analysis: causal conclusions depend on modeling
assumptions beyond a descriptive record alone~\citep{pearl2009}.
The neutral-substrate constraint makes an analogous distinction at
the foundational layer.
A substrate may record that a specified
source asserted a causal or normative proposition without itself
adopting that proposition as fact.
Under the assumptions of
\emph{Neutral Substrates}, the permitted foundational restriction is
sufficient for semantic neutrality~\citep{case2026neutral}.

The present paper examines a different issue: whether identity
judgments about records, rules, occurrences, and their referents
remain justified across transformations.
Annotation, copying,
refinement, and provenance extension may preserve some features and
change others.
Which changes preserve a particular referent's strict
identity is not determined merely by a transformation-family label.
Keeping interpretation attributed also does not establish that the
referential commitments themselves are adequately grounded.

\subsection{FAIR Reuse under Persistent Disagreement}
\label{subsec:related_fair}

The FAIR principles motivate discoverable, accessible, interoperable,
and reusable data across communities and over time~\citep{wilkinson2016}.
They do not, by themselves, establish the referent-specific criteria
needed to justify every co-reference or persistence judgment under
persistent disagreement.
Shared vocabularies, identifiers, schemas,
and documented provenance can support reuse, but their presence alone
does not prove strict identity.

The finite construction studied here complements these concerns by
making selected identity-basis choices and proposed transformation
effects explicit.
Its candidate profiles can serve as comparison
cases where different identity questions arise during reuse.
Whether a particular case calls for separate profiles, an explicit
carrier-indexed representation, or another construction
requires independent analysis.


\section{Limits and Boundary Conditions}
\label{sec:limits}

The results concern a selected finite construction: the reference kinds,
candidate carrier--basis profiles, and transformation basis of
Sections~\ref{sec:formal_setting}--\ref{sec:profiles}.
Within it, the paper derives selected classifications from explicit
models, characterizes invariant-preserving step-generated relations,
and establishes the determinacy of the profile assignment.

\subsection{Conditional Scope}
\label{subsec:limits_conditional_scope}

The analysis is motivated by the neutral-substrate constraint of
\emph{Neutral Substrates}~\citep{case2026neutral} and by the setting of
persistent interpretive disagreement described in
Section~\ref{sec:formal_setting}.
The neutrality result in that earlier work is conditional on Referential
Common Ground and independently justified referential commitments.
Neither the inventory selected here nor its transformation classifications
establishes those premises for a concrete system.

The present results are relative to the selected carrier kinds,
candidate identity bases, eligible transformation instances, and
declared classification assignments.
Their mathematical validity within the specified construction must be
distinguished from the scientific adequacy of those choices.
In particular, a record's stable label, a declared basis, or a
non-breaking transformation classification does not, by itself,
warrant strict identity across presentations.

The certified classification cells establish consequences
of explicit invariant bases, effect models, and warrants.
They do not establish that those models adequately describe
the transformations or identity criteria of an actual system.

Individual derived cells may rely on different presentation
spaces, field decompositions, and admissibility conditions.
Their separate certification does not establish a coherent
common semantic realization of the entire classification matrix.

Empirical counterexamples can challenge a proposed model or
universal verdict.
Finite observations alone do not establish universal conformance
of an unrestricted implementation to that model.

The analysis is not tied to a particular legal theory, policy framework,
or causal account.
That independence of vocabulary does not prove that
its selected identity conditions are valid in every such domain.

\subsection{Completeness and Combination}
\label{subsec:limits_completeness}

The catalog is a selection, not a complete inventory of identity bases
or referent kinds, and its size is a property of the construction.
The single-basis entries do not assert uniqueness.
Records may require distinctions among content, version, token,
signature, and chain of custody.
Occurrences may be analyzed as bounded happenings, phases, or event
complexes.
Obligation-bearing parties may be distinguished from offices, roles,
and organizational units.
Whether any such distinction determines identity depends on
an independently justified criterion for the intended referent.
Extending the transformation basis may reveal further candidate
distinctions or inadequacies in previously assigned classifications;
such outcomes require their own analysis.

The carrier-indexed combination of Section~\ref{sec:distinctions}
shows that separate equivalence relations can be represented together.
Whether a given combination retains every warranted identity judgment
is a question about the identity semantics of the application.

\subsection{Limits of the Selected Setting}
\label{subsec:limits_excluded_systems}

The following cases expose questions that the selected construction
does not settle.
They should not be read as blanket exclusions of
otherwise legitimate representational methods.

\paragraph{Interpretively dependent identity.}
A framework may supply a causal, normative, or institutional criterion
under which a referent is identified.
Such a judgment can be recorded
as an attributed interpretation.
Treating it as an unqualified shared
referential commitment requires independent grounds and compatibility
with the applicable admissible frameworks.
The present profile catalog
does not establish those grounds.

\paragraph{Relation-supported identity.}
Relations may provide evidence or conditions for individuation,
co-reference, or persistence.
Their role cannot be ruled out by
stipulation.
When a construction uses relational criteria, it must
explain what they establish, for which referents, and under what
interpretation conditions.
This paper does not derive a general
prohibition on relation-determined identity.

\paragraph{Framework-relative identity judgments.}
Admissible frameworks may disagree about co-reference or persistence.
Recording their distinct judgments does not itself establish a common
identity judgment.
If the substrate adopts a contested judgment as
foundational, the Referential Common Ground premise must be
re-examined.
Disagreement alone, however, does not prove that all
shared reference has failed.

\paragraph{Declared parameters and undeclared distinctions.}
A role, flag, carrier tag, or identity-basis parameter is not
intrinsically a hidden regime.
An explicit parameter can preserve a
meaningful distinction within a combined representation.
Conversely, an undocumented discriminator or a parameter whose meaning
varies across participating frameworks may fail to support warranted
shared identity judgments.
Whether a proposed combination is faithful
must be assessed rather than decided by its number of labels.

\paragraph{Unmodeled transformations.}
The selected transformation basis $\Fset$ is not complete.
An application may involve operations whose effects differ from the
proposed family-level assignments or require finer typing.
Such cases limit the adequacy of the current catalog for that
application.

\subsection{Scope Exclusions}
\label{subsec:limits_scope_exclusions}

This paper does not construct causal or normative interpretive
frameworks, determine which interpretation should govern, or prescribe
a universal ontology.
It does not provide a method for discovering the
correct identity criterion for an arbitrary referent, certify the
soundness of transformation classifications in deployed systems, or
establish a general procedure for deciding strict identity.

The paper also does not prescribe that every identity-relevant
distinction must be implemented as a separate regime object.
A construction using explicit carrier or basis parameters may be
suitable if its semantics are fixed, its judgments are warranted,
and the required distinctions are preserved.
A context-relative construction may serve a different objective,
provided the limits on shared reference and interpretation are stated.
The scientific questions are what the construction preserves and why
those preservation claims are justified, not merely where a distinction
appears in a schema.

\subsection{Boundary with Accountable Records}
\label{subsec:limits_boundary_accountable_records}

This paper investigates candidate identity bases and their proposed
behavior under transformations of representations.
It does not provide a conformance checker for deployed record systems,
mandate a schema for basis declarations or transformation logs, or
establish that an implementation discloses every identity-relevant
mechanism and use.

Related operational-identity and accountable-records work can examine
whether implemented behavior agrees with a declared criterion and
whether relevant evidence is observable.
Such an audit cannot, solely from agreement with a declaration,
certify that the declaration supplies a scientifically adequate
identity criterion.
Conversely, the construction here does not prescribe implementation
categories for such an audit.

\subsection{Boundary of the Claim}
\label{subsec:limits_boundary_claim}

Within the specified finite construction, the candidate profiles
have pairwise distinct classification functions, the carrier--basis
assignment is determinate, and classification distinctions persist
under extensions that retain the original classifications.
The model-derived comparisons show how different proposed identity
bases assign different effects to an operation family, and the
step-generated results give exact conditions under which their
relations coincide or separate.

These are formal results about declared bases and models.
Whether a basis yields a warranted identity relation for particular
referents depends on independent justification of that basis, the
transformation models, and their warrants.


\section{Conclusion}
\label{sec:conclusion}

This paper pairs candidate identity bases with families of record
transformations and derives, rather than stipulates, how selected
families classify relative to a basis.
Invariant-defined bases and explicit effect models yield verdicts,
certified in Lean~4 for scope decomposition, rule refinement, and
forking, that expose their dependence on field granularity, admissible
steps, and continuity warrants.

The identity relation generated by invariant-preserving steps has an
exact characterization: two bases generate equal relations on a
comparison domain precisely when they agree on the preservation of
every admissible step there, and a single separating step witnesses a
difference.
Under stated assumptions, classification-based relations can be
compared with the step-level construction.
The carrier-indexed combination preserves relations on distinct
carrier kinds within one tagged relation.

Coverage versus internal structure for scopes, content versus
structure for rules, and locus versus object for plain referents serve
as conditional tests of the proposed distinctions, with decomposition,
refinement, and forking as the testing operations.

These results make identity claims about transforming records
auditable, supply testable conditions for systems that transform them,
and identify the assumptions behind each verdict.
They are conditional on the declared bases, effect models, and
warrants.
Naming a reference kind or classifying a transformation does not
settle the identity of its referents, so applying the results to
particular referents requires independently justified individuation,
co-reference, and persistence criteria.

The paper takes its motivation from the conditional neutrality result of
\emph{Neutral Substrates}~\citep{case2026neutral}, which gives
sufficient conditions for semantic neutrality under Referential Common
Ground but does not supply identity criteria for particular referents.

Remaining work is mathematical and scientific: justify the proposed
identity bases for their intended referents, validate transformation
classifications on eligible presentations, relate generated relations
to warranted co-reference and persistence judgments, and determine
which distinctions require separate representation.
The application settings of Section~\ref{sec:application} offer
contexts for these tests, and companion operational-identity and
accountable-records work can examine agreement between implementations
and declared criteria.

Stable reference under persistent disagreement remains the objective.

\section*{Statements and Declarations}

\subsection*{Acknowledgments} The author thanks the anonymous reviewers
of an earlier version of this work for comments that helped improve
the framing, organization, and presentation of the work.

\subsection*{Author Contributions}
The author is the sole contributor to this work and
is responsible for all aspects of the research, authorship, and publication.

\subsection*{Use of AI-Assisted Tools}
AI-assisted tools were used for editing, formatting, and consistency checking,
and in developing and reviewing mathematical definitions, proofs,
and the companion Lean~4 formalization.
The author reviewed all suggestions and is solely responsible for the content.

\subsection*{Declaration of Conflicting Interest}
The author declares no potential conflicts of interest.

\appendix
\clearpage

\section{Pairwise Distinctions and Generated Relations}
\label{app:sep}

The declared classification functions of the candidate profiles in
Table~\ref{tab:master} are different for each pair, as established by
Proposition~\ref{se200.prop.DistinctCoreProfiles}.
This is a result about the \emph{proposed matrix}; distinct
classification functions need not generate distinct identity
relations.

\subsection{Family-Level Generated Relations}

Definition~\ref{se200.def.InducedIdentityRelation} generates
$\sim_\tau$ using admissible steps of families classified
\PRS{} or \NEU{}.

This construction selects steps by their stipulated family
classifications rather than independently testing invariant
preservation at the individual-step level.

Within the stipulated matrix, $\OBL$ and $\REC$ have identical
sets of non-breaking families.
Consequently, on a common eligible domain with the same
admissible transformation dynamics, this paper's construction
generates equal relations for those profiles.
This is a formal comparison; the distinct intended carrier
kinds have not been shown to share a scientifically warranted
comparison domain.

The companion Persistence classification construction differs:
it generates relations using only families licensed \PRS{}.

Within the stipulated matrix, $\OBL/\RULEC$ and
$\SCOPES/\RULES$ have identical sets of \PRS{} families.
They therefore generate identical relations under that
construction on shared transformation dynamics.

Direct enumeration of the stipulated matrix gives seven
distinct \PRS{}-only family sets and eight distinct
\PRS{}-or-\NEU{} family sets.

These counts bound the number of distinct relations generated
on common dynamics by the respective classification-based
constructions.
They do not count scientifically required regimes and do not
bound the invariant-based step relations in general.

All other pairs have different sets of
\PRS{} families and can be separated under suitable dynamics.
Such differences do not prove separation on arbitrary dynamics,
which may lack steps witnessing those differences.

These \PRS-only results do not establish corresponding
equality or separation results for this paper's
\PRS{}-or-\NEU{} generated relation.

\subsection{Invariant-Generated Relations}

The newer invariant-based construction includes the admissible
individual steps that preserve a declared invariant, rather than
selecting entire families solely by their matrix labels.

Theorem~\ref{se200.thm.StepIdentityCollapse} establishes that
two bases on a common effect model and comparison domain
generate equal relations exactly when they agree on invariant
preservation for every admissible step in that domain.

Proposition~\ref{se200.prop.StepSeparation} establishes that
an admissible step preserved by one basis and changed by another
is a sufficient witness of different generated relations,
provided both endpoints lie in the comparison domain.

Conversely, bases that classify a family differently can still
produce equal relations on a restricted domain containing no
separating step.

None of these three generated-relation constructions establishes
strict identity without an independently justified interpretation.

\subsection{Conditional Examples}

The scope, rule, and fork models discussed in
Section~\ref{subsec:model_derived_results} demonstrate how
explicit presentation fields, transformation effects, and
warrants determine derived verdicts.
The fork model is especially instructive: a bare fork does
not entail either a change of locus or a change of object.
The proposed $\LOC/\OBJ$ distinction therefore requires
additional continuity warrants.
Whether the resulting generated relations coincide or differ
also depends on the admissible steps and comparison domain.

The stipulated matrix establishes pairwise differences among
the proposed classification profiles.
These differences do not necessarily imply distinct generated
relations.
The invariant-based step results provide explicit conditions
for relation equality and separation, while their interpretation
as strict identity requires independent scientific justification.

\end{document}